\documentclass[a4paper,11pt]{article}

\usepackage{jinstpub} 
\usepackage{graphicx}
\usepackage{lineno}
\usepackage{todonotes}
\usepackage{textgreek}
\usepackage{fixltx2e}
\usepackage{textcomp}

\title{\boldmath Development of novel low-mass module concepts based on MALTA monolithic pixel sensors}

\author
{
	J Weick$^{1,2}$,
	F Dachs$^{1}$, 
	P Riedler$^{1}$,
	M Vicente Barreto Pinto$^{4}$,
	A M. Zoubir$^{2}$,
	L Flores Sanz de Acedo$^{1}$, 
	I Asensi Tortajada$^{1}$,
	V Dao$^{1}$, 
	D Dobrijevic$^{1,5}$,
	H Pernegger$^{1}$,  
	M Van Rijnbach$^{1,3}$,
	A Sharma$^{1}$, 
	C Solans Sanchez$^{1}$,
	R de Oliveira$^{1}$,
	D Dannheim$^{1}$,
	J V Schmidt$^{1,6}$
}

\affiliation
{
	$^{1}$CERN, Switzerland,
	$^{2}$Technical university of Darmstadt, Germany,
	$^{3}$University of Oslo, Norway,
	$^{4}$Université de Genève, Switzerland,
	$^{5}$University of Zagreb, Croatia,
	$^{6}$Karlsruhe Institute of Technology, Germnany
}

\emailAdd{julian.weick@cern.ch}

\abstract{

The MALTA CMOS monolithic silicon pixel sensors has been developed in the Tower 180 nm CMOS imaging process. It includes an asynchronous readout scheme and complies with the ATLAS inner tracker requirements for the HL-LHC. Several 4-chip MALTA modules have been built using Al wedge wire bonding to demonstrate the direct transfer of data from chip-to-chip and to read out the data of the entire module via one chip only. Novel technologies such as Anisotropic Conductive Films (ACF) and nanowires have been investigated to build a compact module. A lightweight flex with 17 \textmu m trace spacing has been designed, allowing compact packaging with a direct attachment of the chip connection pads to the flex using these interconnection technologies. This contribution shows the current state of our work towards a flexible, low material, dense and reliable packaging and modularization of pixel detectors.
	
}


\proceeding{Topical Workshop on Electronics for Particle Physics\\
	19$^{\text{th}}$-23$^{\text{rd}}$ September\\
	Bergen, Norway}

\begin{document}
	\maketitle
	\flushbottom
\section{Introduction}\label{sec:Intro}	
The upcoming upgrades of the LHC experiments and future detectors are setting new requirements for particle trackers including increased radiation hardness, timing and spacial resolution and rate capability as well as dense integration with a minimal material budget. To realize very large detector surfaces, a scalable production process to manufacture chip modules is also mandatory. This work will show a part of the current efforts of CERNs experimental physics R\&D (EP-R\&D) \cite{RD_report} project to develop flexible, low material, dense, reliable and scalable modules using radiation tolerant pixel chips.
	
\section{Modularization of MALTA}\label{sec:Module}
\begin{wrapfigure}{rh}{8.5cm}
	\includegraphics[width=8.5cm]{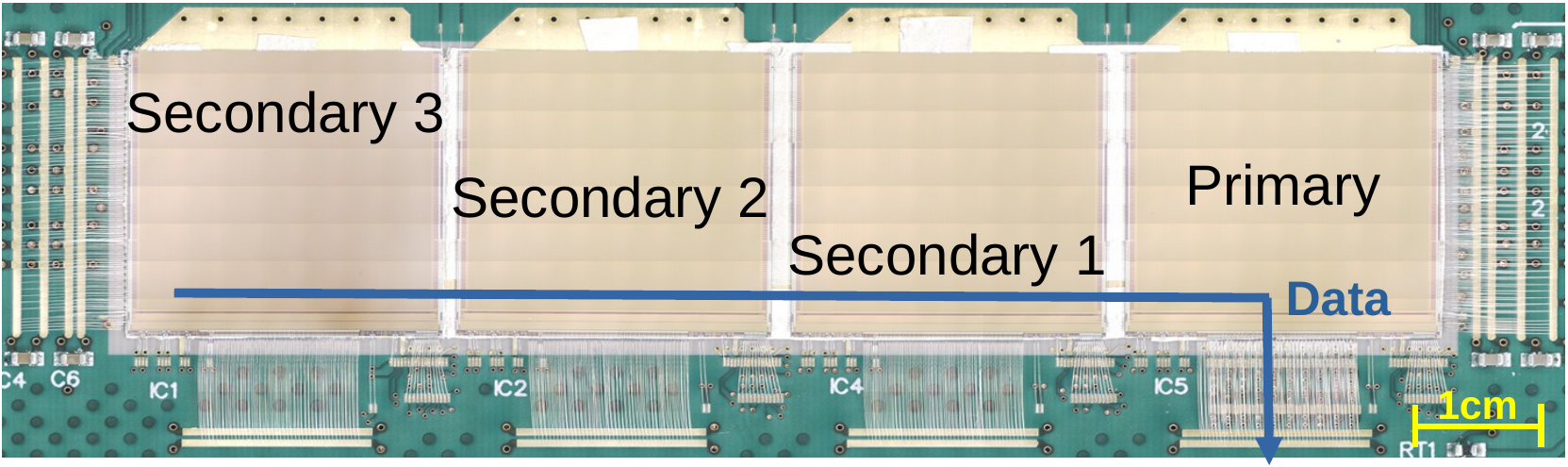}
	\caption{Four chip MALTA module demonstrating the data transfer from secondary (S1-S3) to the primary chip}\label{fig:4chipmodule}
\end{wrapfigure}	
The MALTA chip is a monolithic pixel sensor produced in the 180nm Tower process featuring an asynchronous front-end and read-out. It has proven NIEL radiation hardness up to $3\cdot10^{15}$ n$_{eq}$/cm\textsuperscript{2} , and TID hardness up to 100 Mrad.\cite{b}\\
The data transfer from the secondary (S1-S3) to the primary chip (Prim) has been demonstrated in a 4-chip MALTA module shown in figure \ref{fig:4chipmodule} placed on a rigid PCB carrier housing over one million pixels. The interconnection from chip-to-chip and from chip to PCB is realized using wire-bonds.
Current studies focus on the replacement of wire bonds for chip-to-chip data and power transmission with a silicon interposer (silicon bridge). This allows for a denser integration of the chips, compared to wire bonding, as no minimum spacing between the chips is needed.
Furthermore, a flexible carrier (flex) has been designed to interconnect four MALTA2 chips using a flip-chip \cite{Flip_chip} process, connecting the chip pads to the respective pads on the flex directly.

\section{Interconnection technologies}\label{sec:inteconection_tech}
\begin{wrapfigure}{rh}{7.5cm}
	\includegraphics[width=7cm]{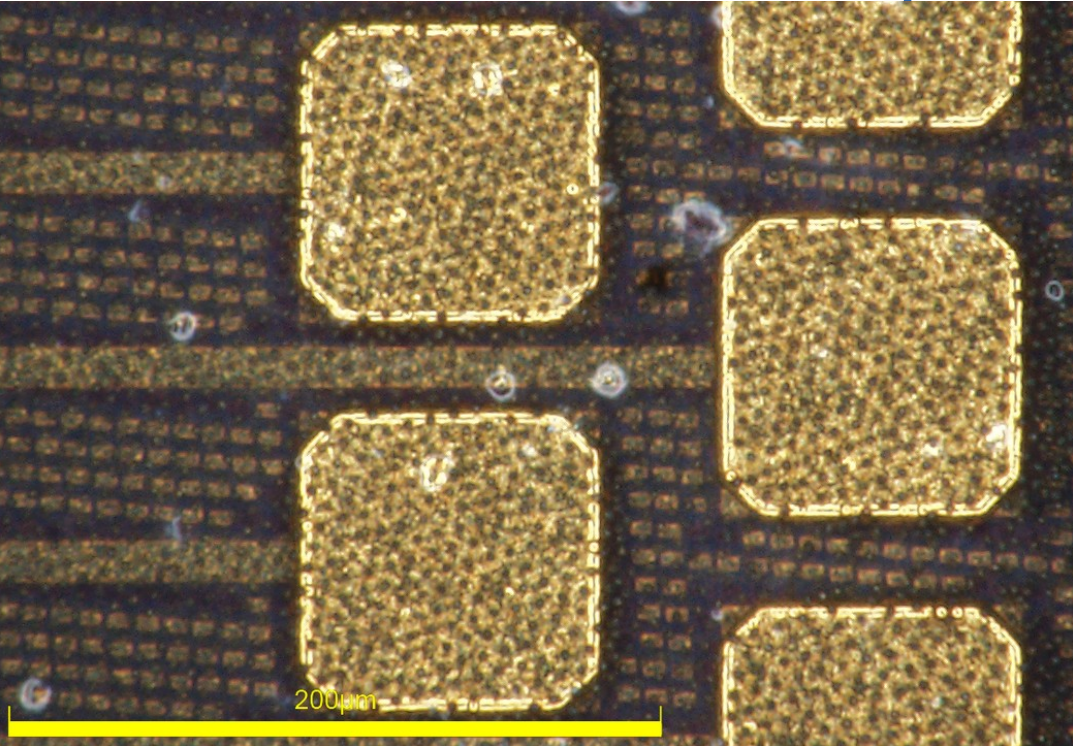}
	\caption{ACF (black dots) on ENIG plated MALTA pad}\label{fig:ACF_on_MALTA_pad}
\end{wrapfigure}
The small pad size of MALTA and its dense pad layout require the study of new interconnection technologies such as anisotropic conductive films (ACF) or a surface nano structuration process (nanowires).
These technologies have the potential to provide a scalable chip-to-chip or chip-to-flex interconnection realizable in a fast interconnection process and are suitable for a large number of pads. The interconnection is demonstrated using the aluminum pads of size 88$\times$88~\textmu m$^2$ of the MALTA chip.

\subsection{Anisotropic conductive film (ACF)}

ACF is a industry standard interconnection technology used in LCD screen production \cite{LCD}. The interconnection is established with metal-coated polymer particles, embedded in glue. Figure \ref{fig:ACF_on_MALTA_pad} shows the ACF balls (black dots) on the MALTA pads.
In a preparatory step, the Aluminium pads are plated using Electroless Nickel Immersion Gold (ENIG) \cite{a}, which serves as an elevation layer for the ACF particles and prevents the formation of an oxide layer. 
Afterwards, the device to be bonded is laminated with the ACF. Finally, the assembly is done in a flip-chip process using pressure to compress the ACF conductive particles between pads and heat to cure the glue. As such, ACF provides a cost-effective, maskless, and in-house capable assembly technology.
ACF is currently developed as a bonding process for the silicon interposer for MALTA modules, and will be used in the flip chip assembly of the MALTA2 chips on the flex carrier introduced in section \ref{sec:flex}. 

\subsection{Nanowires}

\begin{wrapfigure}{rh}{8.4cm}
	\includegraphics[width=7.5cm]{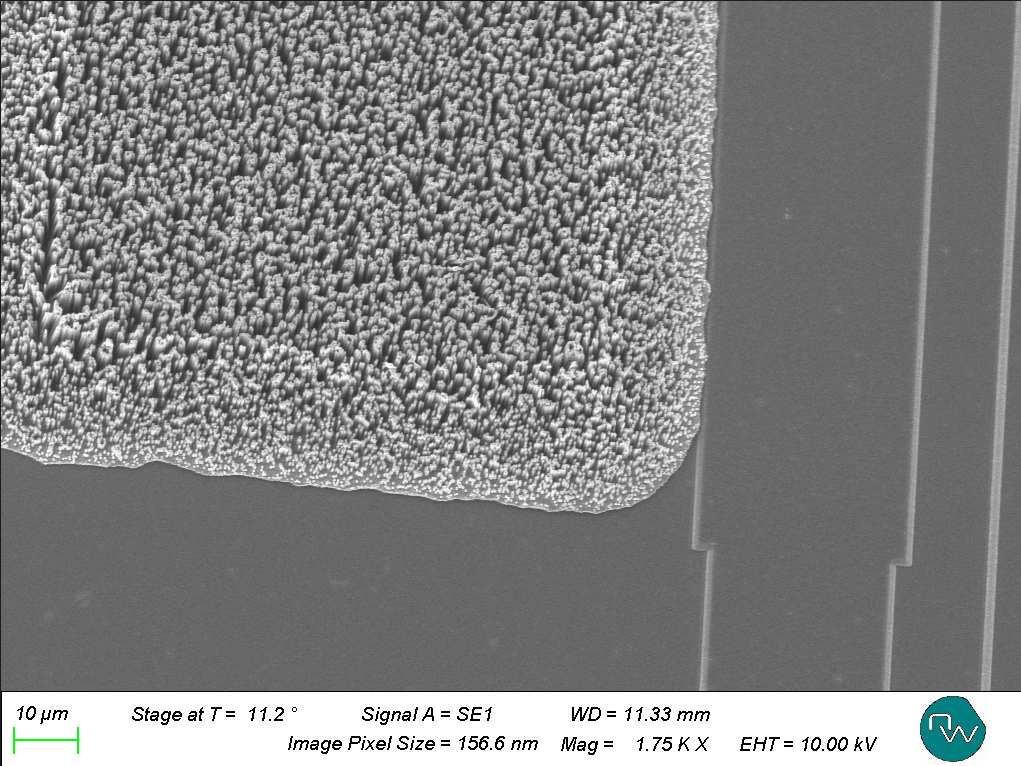}
	\caption{Copper nanowires grown on 88$\times$88~\textmu m$^2$ MALTA pad}\label{fig:Wires_on_MALTA_pad}
\end{wrapfigure}

Nanowires are evaluated as a second interconnection technology. For this bonding process, a titanium layer is deposited on the aluminum pads of MALTA together with a gold finish to prevent oxidation.
This serves as a base for a copper seed layer on which nanowires with a diameter in the hundred-nanometer range, and a length of several micrometers are grown. Figure \ref{fig:Wires_on_MALTA_pad} shows the copper nanowires on a MALTA pad.\\
The bonding process can be realized using three different procedures, which differs in their applied pressure and heat, and the necessity to apply the nanowires on only one side (chip pads or targeted carrier pads), or on both:\\
\begin{itemize}
	\item \textbf{Sinter process:} The wires are applied to one side only. The connection is formed via a sintering process that requires high pressure and heat, compared to the other procedures.
	\item \textbf{Cold welding process:} Wires are applied on both sides and allow for a large contact surface. The connection is established with a cold-welding process that requires minimal pressure and heat.
	\item \textbf{Gluing process:} In this bonding process nanowires are only applied on one side. A non-conductive glue is used as underfill which allows for reduced pressure and heat and enhanced mechanical stability.
\end{itemize}
Nanowires promise to offer low contact resistance and parasitic inductance/capacitance, and can be applied either on the targeted carrier or directly on the chip (chip or wafer level).

\section{Interconnection test structure}\label{sec:test_structure}

A test structure has been developed on a 300 \textmu m aluminum nitride (AlN) ceramic to validate the introduced interconnection technologies electrically and mechanically as well as pre-optimizing the process parameters for interconnecting the MALTA2 chips with the flex presented in section \ref{sec:flex}.\\

\begin{figure}[h]
\centering
	\includegraphics[width=8.5cm]{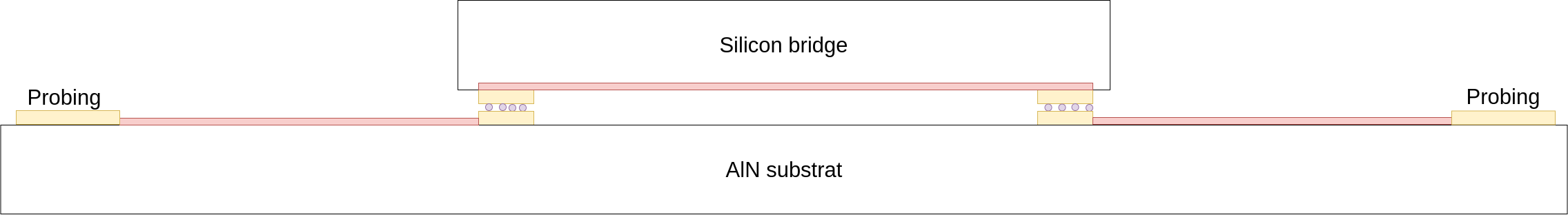}
	\caption{Concept of the silicon bridge test setup, to test the electrical characteristic of two ACF interconnections}\label{fig:HF_setup}
\end{figure}

\begin{figure}[h]
	\centering
		\begin{minipage}{.4\linewidth}
		\centering
		\includegraphics[width=4.5cm]{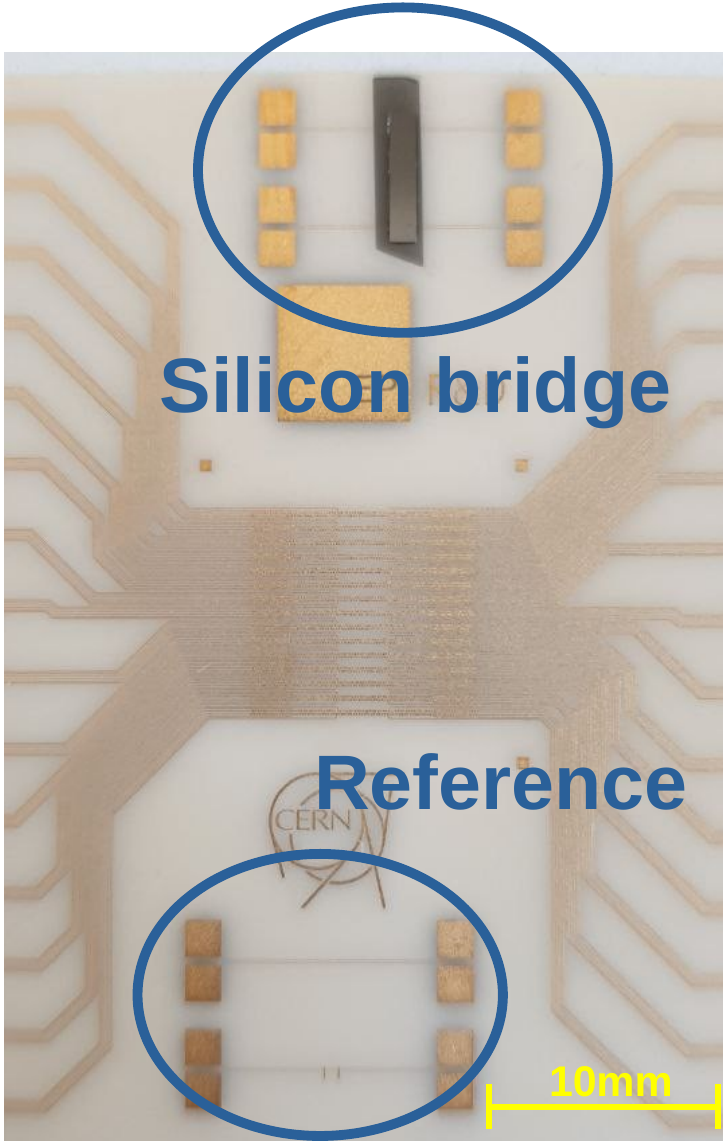}
		\caption{ACF bonded silicon MALTA interposer and reference on test structure for electrical characterization.}
		\label{fig:Test_structure}
	\end{minipage}
	\hfill
	\begin{minipage}{.55\linewidth}
		\centering
		\includegraphics[width=7.7cm]{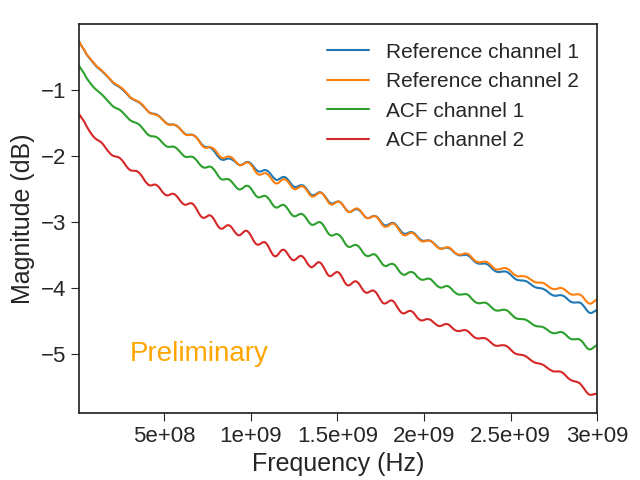}
		\caption{Signal dampening from 0 to 3 GHz over a ACF bonded silicon bridge and reference structure}
		\label{fig:HF_test_structure}
	\end{minipage}
\end{figure}
AlN has been chosen due to its similar thermal conductivity to silicon and compatibility with the manufacturing process of the flex. The various pad layouts of this structure mirror the pads of the MALTA chip and the silicon interposer. They are designed to study, among other parameters, the bond success rate per pad, the probability of electrical shorts between neighboring pads and the resistivity of the interconnection.
First electrical tests indicate a DC resistance between 1.7 \textOmega\, to 3 \textOmega\, for the ACF interconnection on the 88$\times$88 \textmu m\textsuperscript{2} MALTA pad using an ACF with 3 \textmu m Ni-Au/Polymer conductive particles, while the current AlSi wire bonds with a diameter of 25\textmu m feature a resistance of <1 \textOmega\ on our current carriers \cite{Wire_bond}. The setup to test the insertion loss of a differential signal over an ACF connection compared to a directly wired reference is shown in figure \ref{fig:HF_setup}. The silicon bridge is bonded onto the test structure using ACF, and establishes a electrical connection between the probing pads. Figure \ref{fig:Test_structure} shows the bonded bridge on the substrate with the silicon bridge in top and the reference structure on the bottom.\\
Since the structure is optimized for mechanical test rather than for high frequencies (HF) it only allows for a qualitative statement of the signal decay compared to the reference. The signal decay has been evaluated using a vector network analyzer with Picotest probes which allow to probe the large probing pads directly. First results are shown in figure \ref{fig:HF_test_structure}.\\
The signal magnitude over the reference as well as over the ACF interconnection decays at higher frequencies. The measured offset of the ACF connection in relation to the reference (<2dB) is indicating an acceptable signal loss for our purposes.The difference between ACF channel 1 and ACF channel 2 is in line with the spread in DC resistance between different pads and a result of the non optimized bonding process. To validate the signal transfer over ACF, further investigation in combination with a dedicated HF test structure is needed.

\section{Low mass flexible MALTA2 carrier} \label{sec:flex}
\begin{wrapfigure}{rh}{7.5cm}
	\includegraphics[width=7.5cm]{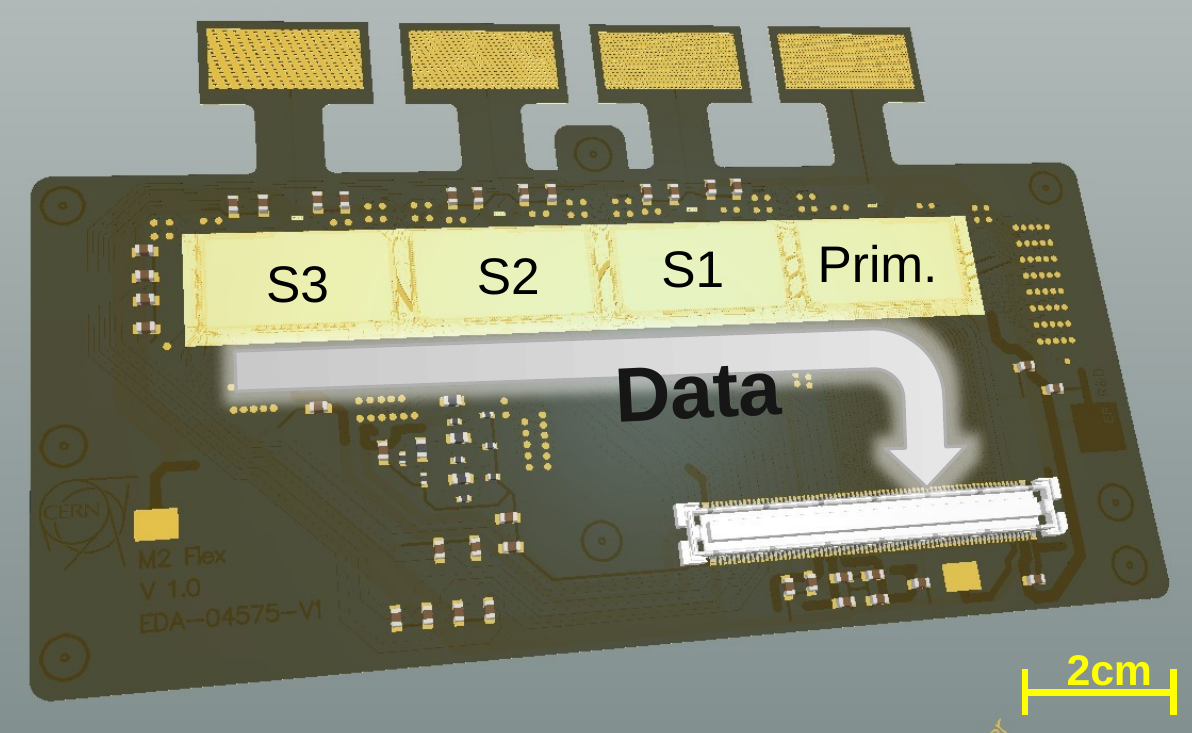}
	\caption{Flex carrier for four MALTA2 chips with chip to chip data transfer}
	\label{fig:Flex}
\end{wrapfigure}

A Low mass flexible MALTA2 carrier (flex) for a 4-chip module is currently produced as a proof-of-concept for flip-chip mounting the MALTA2 chip on a flexible support structure, using minimal material while providing maximal testing capabilities.
Figure \ref{fig:Flex} shows a model of the first flex prototype. We use the data readout scheme validated with the four-chip MALTA PCB in section \ref{sec:Module}, transferring the data from three secondary chips (S1-S3) to a primary chip (Prim) and from there over a dedicated connector flex circuit to the data acquisition FPGA. The layout of the flex allows for the chips to be tuned and powered individually. This gives us the possibility to test the influence of constrained powering on the performance and the data transfer capabilities of the module.\\
Due to the pad layout of MALTA, wire bonds only allow for arranging the chips in a single row module. This constraint does not apply to modules assembled on a flex circuit using a flip chip process. Furthermore, chips can be supplied with power through the flex support structure allowing to increase the module size compared to chip-to-chip powering. The same holds true if the tiling is replaced by a wafer-scale integrated sensor using stitched devices.\\
\begin{wrapfigure}{lh}{8.5cm}
	\centering 
	\includegraphics[width=4cm]{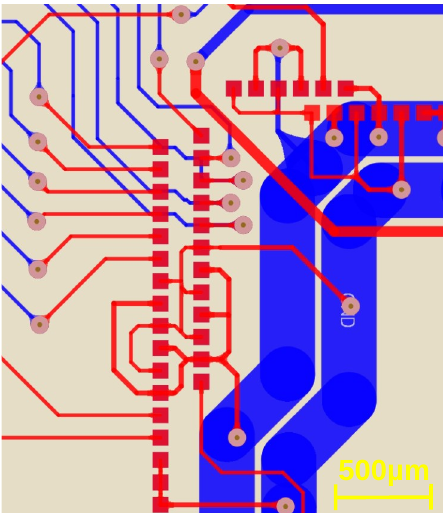}
	\includegraphics[width=4cm]{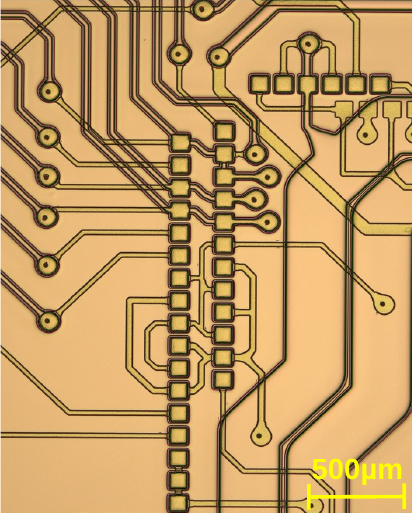}
	\caption{Layout (left) and produced flex (right)}
	\label{fig:Flex_and_layout}
\end{wrapfigure} 
The dense pad layout of the MALTA chip requires a flex circuit with fine structures in order to keep capacitive loads low and minimize the number of metal layers and thus material budget.\\
The chosen flex production technology allows for structure sizes down to 15 \textmu m track and clearance, a 10 \textmu m polyamide layer thickness, and a 6 \textmu m copper thickness. Together with 20 \textmu m solder stop, the designed flex circuit has an overall thickness of roughly 50 \textmu m. The 20 \textmu m solder stop is only needed for assembly and population of the circuit as well as to enable soldering on debug pads. Figure \ref{fig:Flex_and_layout} shows the manufactured flex on the right side and the respective layout on the left.

\section{Summary and outlook}
	
In this work, we present first results from studies on mechanically robust and scalable interconnection techniques, which could offer an alternative to wire-bonding.\\
We show an approach to densely package multiple MALTA pixel chips onto a flexible module carrier minimizing the needed material. In a next step, the flex will be assembled using the introduced interconnection technologies, followed by tests in the lab and at test beam to quantify the performance of the module. \\

\end{document}